\documentclass[useAMS,usenatbib,usegraphicx]{mn2e}

\def\msun{M$_{\sun}$}
\def\rsun{R$_{\sun}$}
\def\muhz{$\mu$Hz}
\def\aap{A\&A}
\def\apjl{ApJ}
\def\apj{ApJ}

\def\mnras{MNRAS}
\def\nat{Nature}
\def\pasp{PASP}

\newcommand {\lessim} {\ {\raise-.5ex\hbox{$\buildrel<\over\sim$}}\ }

\title[Pulsations in PSR 1738+0333B]
{PSR J1738+0333: The First Millisecond Pulsar + Pulsating White Dwarf Binary\thanks{
Based on observations obtained at the Gemini Observatory, which is operated by the 
Association of Universities for Research in Astronomy, Inc., under a cooperative agreement 
with the NSF on behalf of the Gemini partnership: the National Science Foundation 
(United States), the National Research Council (Canada), CONICYT (Chile), the Australian 
Research Council (Australia), Minist\'{e}rio da Ci\^{e}ncia, Tecnologia e Inova\c{c}\~{a}o 
(Brazil) and Ministerio de Ciencia, Tecnolog\'{i}a e Innovaci\'{o}n Productiva (Argentina).
}
}

\author[M. Kilic et al.]
       {Mukremin Kilic$^{1}$,
       J. J. Hermes$^2$,
       A. Gianninas$^1$,
       Warren R. Brown$^3$\\
       $^1$Department of Physics and Astronomy, University of Oklahoma, 440 W. Brooks St., Norman, OK, 73019, USA\\
       $^2$Department of Physics, University of Warwick, Coventry CV4 7AL, UK\\
       $^3$Smithsonian Astrophysical Observatory, 60 Garden St, Cambridge, MA 02138, USA\\
}

\begin{document}

\maketitle

\begin{abstract}

We report the discovery of the first millisecond pulsar with a pulsating white dwarf companion. Following the recent discoveries of pulsations in extremely low-mass (ELM, $\leq0.3$ \msun) white dwarfs (WDs), we targeted ELM WD companions to two millisecond pulsars with high-speed Gemini photometry. We find significant optical variability in PSR~J1738+0333 with periods between roughly $1790-3060$~s, consistent in timescale with theoretical and empirical observations of pulsations in $\approx$0.17 \msun\ He-core ELM WDs. We additionally put stringent limits on a lack of variability in PSR J1909$-$3744, showing this ELM WD is not variable to $<0.1$ per cent amplitude. Thanks to the accurate distance and radius estimates from radio timing measurements, PSR~J1738+0333 becomes a benchmark for low-mass, pulsating WDs. Future, more extensive time-series photometry of this system offers an unprecedented opportunity to constrain the physical parameters (including the cooling age) and interior structure of this ELM WD, and in turn, the mass and spin-down age of its pulsar companion. 

\end{abstract}


\section{INTRODUCTION}

Optical counterparts to radio millisecond pulsars (MSP) provide exceptional constraints on the masses and thus equations of state of their neutron star companions. These optical counterparts are very commonly low-mass white dwarfs (WDs), whose progenitors are responsible for spinning up the pulsars to millisecond rotation rates (see \citealt{vankerkwijk05,tauris12,smedley14} and references therein).

Exploiting these unique WD+MSP systems often requires detailed knowledge of extremely low-mass (ELM, $\leq 0.3$ \msun), He-core WDs. For example, \citet{antoniadis13} constrain the mass of PSR J0348+0432 to $2.01 \pm 0.04$ \msun, and thus use it as a probe for strong-field gravity, by exploiting optical observations of the 0.172~\msun\ ELM WD in the system.

Importantly, deriving a WD mass from its atmospheric parameters is highly sensitive to the assumed mass of the outer hydrogen layer. There is a heretofore observationally unconstrained boundary above which ELM WDs suffer at least one or a series of diffusion-induced CNO flashes leading to thin hydrogen envelopes. Theoretical models suggest this transition occurs between roughly 0.17 to 0.22 \msun\ (e.g., \citealt{driebe98,panei07,althaus13}).

Empirical tests to this scenario by way of constraints on the mass-radius relationship for low-mass WDs are still sparse, but growing thanks to the number of eclipsing and tidally distorted ELM WDs known (e.g., \citealt{steinfadt10a,brown11,hermes14}). However, \citet{kaplan14} show that there is significant (up to 25 per cent) disagreement between the spectroscopically determined mass of the eclipsing ELM WD NLTT 11748 and the mass derived from detailed light curve modelling.

Fortunately we have an avenue into the interiors of these stars through asteroseismology. As WDs cool, they develop partial-ionization zones in their non-degenerate atmospheres that depends on the dominant photospheric composition, which effectively drive global stellar oscillations. For hydrogen-dominated WDs, this onset occurs when $T_{\rm eff}\leq12\,500$~K: these are the DAVs (or ZZ Ceti stars). Pulsation periods are typically anti-correlated with surface temperature, with larger pulsation periods seen for cooler DAVs (e.g., \citealt{mukadam06}). The non-radial $g$-mode pulsations in pulsating WDs have been successfully used to probe the overall mass, rotation rate, convective efficiency, and, perhaps most importantly, the hydrogen envelope mass of these stars \citep{winget08,fontaine08,althaus10}.

\citet{steinfadt10} were the first group to explore the pulsational instability of low-mass ($M<0.45$\msun) WDs, demonstrating that such objects should pulsate at lower temperatures (compared to the normal DAVs), with longer periods and larger mean period spacings. Subsequent to a detailed but unsuccessful search for pulsations in 12 low-mass WDs by \citet{steinfadt12}, the first five pulsating low-mass WDs have since been discovered \citep{hermes12,hermes13a,hermes13b}. These have $T_{\rm eff}= 8000 - 10,000$ K and $\log{g}= 6 - 7$ cm s$^{-2}$ \citep{gianninas14}, which correspond to $0.16 - 0.23$ \msun, and show multi-mode pulsations with periods ranging from about 1200~s to 6200~s.

\citet{corsico12} and \citet{vangrootel13} presented a non-adiabatic stability analysis of low-mass WDs, and both predict unstable $g$-modes with pulsation periods of $\geq$500-1100 s, and up to $\approx$8000 s. \citet{vangrootel13} conclude that the instability strip observed for low-mass WDs is a natural extension of the ZZ Ceti instability strip to lower surface gravities and cooler temperatures. Based on the empirical boundaries of the instability strip presented in \citet{gianninas14}, we identify the ELM WD companions to PSR~J1738+0333 and PSR~J1909$-$3744 as strong candidates for pulsation instability.

In this Letter, we provide time-series observations of these two targets and report the discovery of pulsations in PSR~J1738+0333. Our observations are discussed in Section 2, and the light curves for PSR~J1909$-$3744 and PSR~J1738+0333 are presented in Sections 3 and 4, respectively. We conclude with a discussion of the instability strip for ELM WDs and the future prospects for studying this unique binary system in Section 5.

\section{OBSERVATIONS}

We obtained time-series photometry of PSR~J1909$-$3744 using the 8m Gemini-South telescope with the Gemini Multi-Object Spectrograph (GMOS) on UT 2013 Sep 29 as part of the queue program GS-2013B-Q-53. We obtained 246 $\times$ 40.5 s exposures through an SDSS-$g$ filter over 3.8 h. To reduce the read-out time and telescope overhead to $\approx$15 s, we binned the chip by 4$\times$4, which resulted in a plate scale of 0.29 arcsec pixel$^{-1}$. Observations were obtained under photometric conditions with a median seeing of 0.9 arcsec.

We obtained time-series photometry of PSR~J1738+0333 using the same setup as above, but on the 8m Gemini-North telescope on UT 2014 June 23 as part of the queue program GN-2014A-Q-23. We obtained 243 $\times$ 50 s exposures using an SDSS-$g$ filter over 5.5 h. Our observations were interrupted by a target-of-opportunity program for about an hour, causing a gap in our coverage of this field. Conditions were photometric with a median seeing of 0.5 arcsec.

For reductions and calibrations, we use the standard Gemini GMOS routines under Image Reduction and Analysis Facility supplied with the daily bias and twilight sky frames. For GMOS South (North), we identify 44 (17) non-variable reference stars that are on the same CCD and amplifier as the target, and use them to calibrate the differential photometry. The r.m.s. scatter in our differential photometry of the low-mass WDs in PSR~J1909$-$3744 ($V \approx 21$ mag) and PSR~J1738+0333 ($V = 21.3$ mag) are $<0.02$\% and $<0.2$\%, respectively.

\section{PSR J1909$-$3744}

\begin{figure}
\vspace{-0.2in}
\includegraphics[width=0.991\columnwidth]{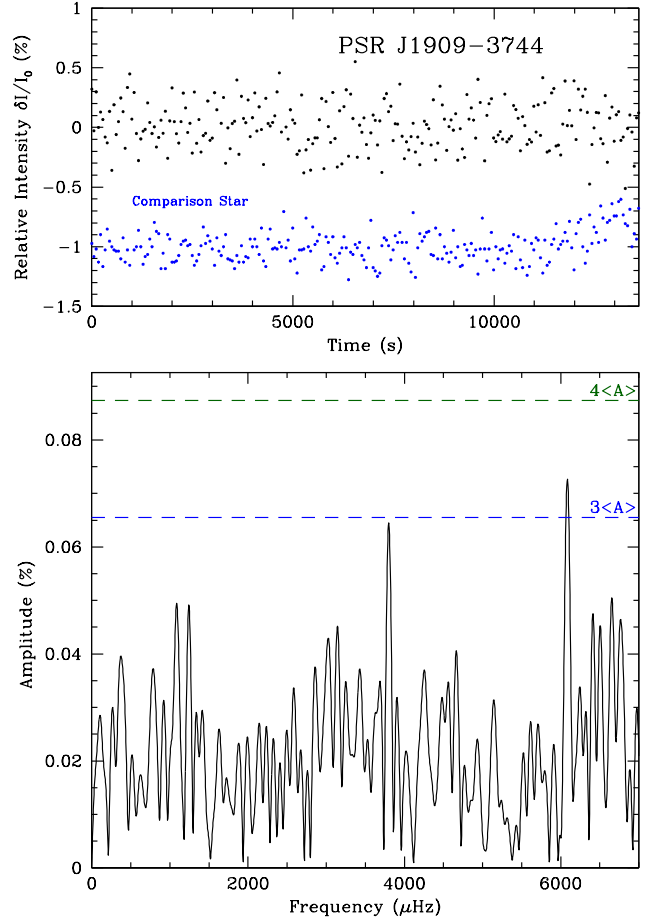}
\caption{The Gemini-South light curve (top panel) and FT (bottom panel) for the optical companion to PSR~J1909$-$3744. For reference, a comparison star is shown in blue, offset by $-1$\%. We mark the 4$\langle {\rm A}\rangle$ and 3$\langle {\rm A}\rangle$ significance level, described in the text, as dashed green and blue lines, respectively.
\label{fig:psr1909}}
\end{figure}

PSR~J1909$-$3744 is a 2.95~ms pulsar in a 1.53~d, nearly edge-on orbit with a $T_{\rm eff} \sim 8500$~K WD at a distance of 1.14~kpc. The Shapiro delay measurements indicate masses of $M = 0.2038 \pm 0.0022$ \msun\ for the WD and $M = 1.438 \pm 0.024$ \msun\ for the pulsar \citep{jacoby03,jacoby05}.

Figure~\ref{fig:psr1909} shows the Gemini light curve and its FT for the ELM WD companion to PSR~J1909$-$3744. There are no significant variations down to a 4$\langle {\rm A}\rangle$ limit of 0.09 per cent relative amplitude for this WD, where $\langle {\rm A}\rangle$ is the mean amplitude of the FT from $0-9000$ \muhz. Hence, our data firmly rule out pulsations in this system. Given the accurate mass estimate for the WD companion from the radio data, ruling out variability in PSR~J1909$-$3744 helps put excellent constraints on the instability strip for low-mass, He-core WDs, as discussed in Section~\ref{sec:istrip}.

\section{PSR J1738+0333}

PSR~J1738+0333 is a 5.85~ms pulsar in a 8.5~h orbit with a $T_{\rm eff}= 9130 \pm 150$ K, $\log{g}= 6.55 \pm 0.07$, $M=0.181^{+0.007}_{-0.005}$\msun, and $R\approx0.04$\rsun\ WD at a distance of 1.47~kpc \citep{antoniadis12,freire12}. The radial velocity observations of the ELM WD and the pulsar timing measurements imply that the pulsar has a mass of $1.47^{+0.07}_{-0.06}$\msun\ and the inclination of the binary is $i=32.6^{\circ}\pm1.0^{\circ}$. The orbital decay of the binary, $dP/dt = -25.9 \pm 3.2 \times 10^{-15}$ s s$^{-1}$, is consistent with the predictions from General Relativity to within $1\sigma$ \citep{freire12}.

\begin{figure}
\vspace{-0.2in}
\centering{\includegraphics[width=0.991\columnwidth]{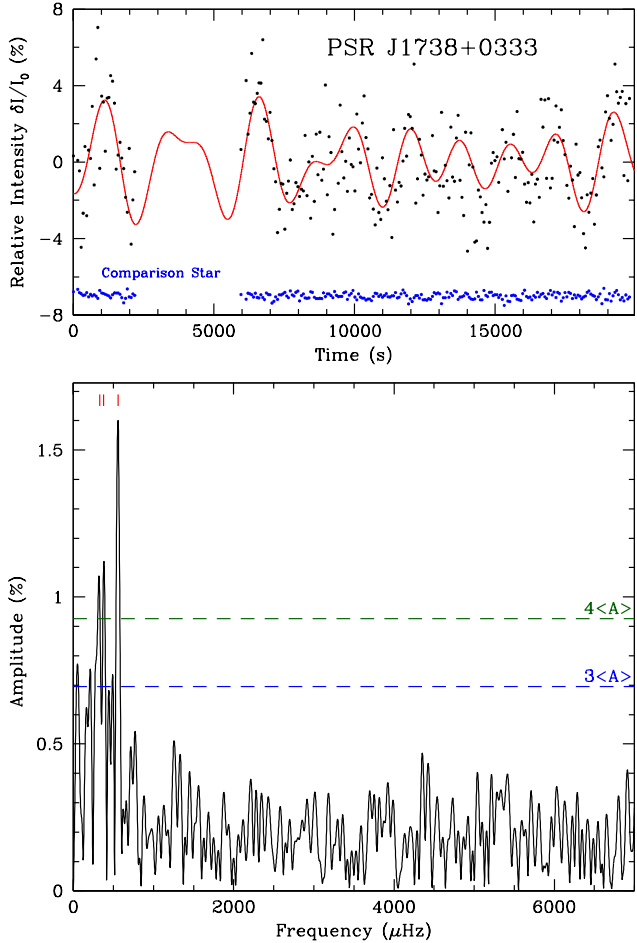}}
\caption{The GMOS-N light curve (top panel) and FT (bottom panel) for the optical companion to PSR~J1738+0333. As with Figure~\ref{fig:psr1909}, a comparison star is shown in blue, offset by $-7$\%. We mark the 4$\langle {\rm A}\rangle$ and 3$\langle {\rm A}\rangle$ significance level, described in the text, as dashed green and blue lines, respectively. The three significant pulsations are marked with red lines in the bottom panel and the frequency solution detailed in Table~\ref{tab:J1738freq} is illustrated in the top panel.  \label{fig:psr1738}}
\end{figure}

Figure~\ref{fig:psr1738} shows the Gemini light curve and its FT for the WD companion to PSR~J1738+0333. The ELM WD companion in this system is pulsating with at least three significant periods of variability, becoming the first milli-second pulsar + pulsating WD binary known.

The FT shows a dominant peak at $\approx$1800 s. Fitting only this highest-amplitude signal (single-mode 
solution), we find it has a 1.60 per cent amplitude. However, we also see evidence for two additional
significant periods of variability at 3057 and 2656 s.

Table~\ref{tab:J1738freq} presents the multi-mode frequency solution from a multiple nonlinear least-squares fit to the three highest peaks in the FT. The frequency, amplitude, and phase uncertainties come from the 3$\sigma$ least-squares values. We also include the signal-to-noise ratio of each signal, calculated by comparing the mode amplitude to the average FT amplitude after subtracting out the three significant periods of variability.

Interestingly, \citet{antoniadis12} noted that there is a relatively large scatter, of order 0.05 mag, in their acquisition frames for the spectroscopic observations of the companion to PSR~J1738+0333 that was not correlated with orbital phase. This is, in hindsight, likely the first evidence of pulsations in the system.

Optical variability in pulsar companions can also be due to irradiation from the pulsar.
In fact, there are about a dozen binary systems in which the high energy emission from
the pulsar heats the atmosphere of the companion and gives rise to optical variability. The black
widow pulsar PSR B1957+20 is an excellent example of such a system \citep{vanparadijs88,fruchter88}.
Assuming a spin-down luminosity of $4.8 \times 10^{33}$ ergs s$^{-1}$ and an irradiation
temperature of 3800 K, the expected orbital modulation in PSR J1738+0333 is only $\Delta L / L = 0.4$\%
\citep[see][]{antoniadis12}. This is below the 3$\langle {\rm A}\rangle$ detection limit of our observations.
Hence, irradiation by the pulsar cannot explain the observed optical variability in PSR J1738+0333's WD
companion.

%TABLE1
\begin{table}
\centering
\caption{Multi-mode frequency solutions for the ELM WD companion to PSR J1738+0333 \label{tab:J1738freq}}
\begin{tabular}{ccccc}
\hline
Period & Frequency & Amplitude & Phase & S/N\\
 (s)   & ($\mu$Hz)    &   (per cent)    &  (s) &    \\
\hline
1788 $\pm$ 33 & 559 $\pm$ 10 & 1.27 $\pm$ 0.47 & 1030 $\pm$ 110 & 6.3 \\
3057 $\pm$ 99 & 327 $\pm$ 11 & 1.22 $\pm$ 0.47 & 10 $\pm$ 190 & 6.0 \\
2656 $\pm$ 80 & 376 $\pm$ 11 & 1.15 $\pm$ 0.47 & 2150 $\pm$ 170 & 5.7 \\
\hline
\end{tabular}
\end{table}

At 9130~K, the WD companion to PSR~J1738+0333 is only slightly cooler than the 9380~K pulsating ELM WD J1840+6423 \citep{hermes12, gianninas14}; both have a surface gravity of $\log{g}= 6.55$ determined from model atmosphere fits to the spectroscopic observations. J1840+6423 shows pulsations at multiple significant periods between $2094-4890$ s \citep{hermes13b}; this range is qualitatively similar to the optical variability observed in the WD companion to PSR~J1738+0333.

\citet{corsico12} performed a non-adiabatic analysis of pulsation instabilities in a $M=0.17$\msun\ He-core WD, and found that $g$-modes become unstable and likely reach observable amplitudes at $T_{\rm eff} \approx 9200$ K. The unstable oscillations in their models have radial order $k \geq 9$, with periods $\geq$ 1100 s for $\ell=1$ modes. At $T_{\rm eff} = 9130 $ K, they predict $g$-mode pulsations with periods ranging from 1100~s to $\approx$ 3000~s (see their Fig. 15), exactly in line with the optical variability we see in PSR~J1738+0333.

\section{DISCUSSION AND CONCLUSIONS}

\subsection{A Benchmark Pulsating ELM WD}

With the discovery of pulsations in the ELM WD companion of PSR~J1738+0333, there are now six pulsating ELM WDs known \citep{hermes13b}. None of the other known pulsating ELM WDs have parallax measurements available. Hence, the companion to PSR~J1738+0333 is the only pulsating ELM WD with independent and precise distance and radius measurements. It becomes a benchmark for pulsating ELM WDs and the modeling of pulsations in He-core WDs.

\citet{antoniadis12} derived the mass of PSR~J1738+0333 using the mass-radius relations of \citet{panei00}, and found it to be roughly 0.18 \msun. This places it right on the boundary of ELM WDs that undergo CNO flashes, which can significantly alter the resultant hydrogen layer mass and thus the cooling rate \citep{althaus13}. A detailed asteroseismic analysis of the pulsations in PSR~J1738+0333 can empirically test whether this ELM WD has a thick hydrogen layer, and better constrain its overall mass.

\citet{corsico12} and \citet{vangrootel13} demonstrate that the period spacing between consecutive radial modes strongly depend on the stellar mass, as well as the thickness of the outer hydrogen envelope. The mean period spacing between adjacent modes is significantly smaller for the thick envelopes compared to the thin envelopes. If a rich spectrum of periods are observed, both the stellar mass and the thickness of the surface H layer can be constrained. The latter has a significant impact on the amount of residual H burning and the cooling rate of the WD. This can be used to put a lower limit on the cooling age of the WD and therefore, the spin-down age of its pulsar companion. 

In addition, the non-radial pulsations in the ELM WD companion to PSR~J1738+0333 offer another astrophysical clock to this richly timed system. While it is not as sensitive as radial-velocity observations, we expect a 5.56 s light-travel modulation in the pulsation arrival times of this ELM WD every 8.5~hr as the massive MSP moves the WD around a common center of mass. We have not included this correction in our preliminary frequency solution, since it is well below the 1$\sigma$ least-squares phase uncertainties.

\subsection{The ELM WD Instability Strip}
\label{sec:istrip}

The addition of two more pulsar companions in this paper brings the total sample size to four WD+MSP systems with optical time-series photometry. Previously,
\citet{steinfadt10} identified three candidates, including the companion to PSR~J1012+5307, for follow-up time-series photometry based on their initial models for pulsating ELM WDs. They also predicted that the companion to PSR~J1911$-$5958A is likely too hot to pulsate. \citet{steinfadt12} observed 12 low-mass WDs, and showed that the companions to these two pulsars do not pulsate down to detection limits of 2.0 per cent and 1.6 per cent, respectively. \citet{hermes13b} put more stringent limits on the lack of variability for~PSR J1012+5307, and showed that it is stable to 0.7 per cent. 

\begin{figure}
\vspace{-0.6in}
\includegraphics[width=3.4in,angle=0]{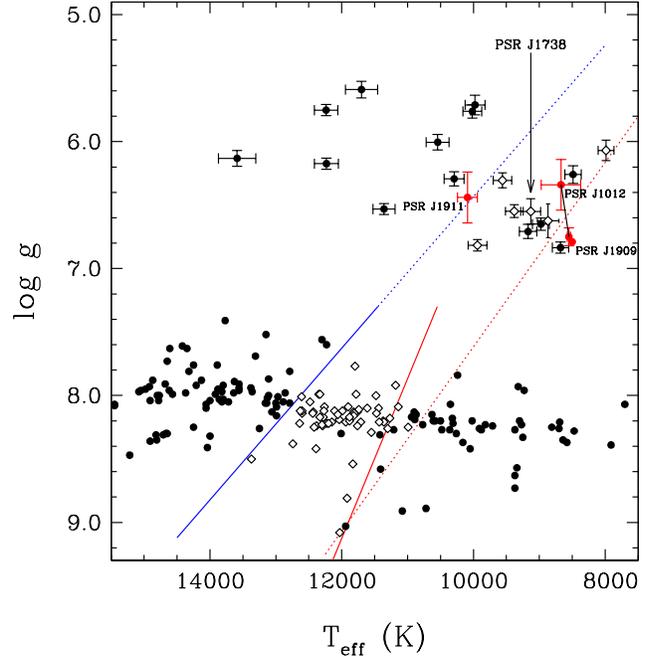}
\vspace{-0.5in}
\caption{The ZZ Ceti instability strip, including the pulsating (open diamonds) and non-pulsating (filled circles) WDs. The pulsating WD companion to PSR~J1738+0333 is labeled. The companions to the other three pulsars are shown as red, filled circles. The WD companion to PSR~J1012+5307 has two temperature and surface gravity measurements in the literature; those are connected by a solid line. The blue and red solid lines represent the empirical ZZ Ceti instability strip from \citet{gianninas11}, whereas the dotted lines show the tentative boundaries from \citet{gianninas14}.
\label{fig:strip}}
\end{figure}

Figure \ref{fig:strip} presents the instability strip for the normal ZZ Ceti stars \citep{gianninas11} and the six currently known pulsating ELM WDs, including the companion to PSR~J1738+0333. The surface temperature and gravity measurements for all objects come from a model atmosphere analysis using the same set of model grids \citep{gianninas14}. In contrast, the atmospheric parameters for the ELM WD companions to the four pulsars discussed above come from a variety of sources \citep{vankerkwijk96,callanan98,bassa06}. Hence, we do not try to update the empirical boundaries of the current instability strip, but instead we provide a comparison of those parameters against the known pulsators. 

PSR~J1738+0333 falls right in the middle of the empirical instability strip \citep[as outlined in][]{gianninas14}, though its parameters are also similar to two WDs that do not pulsate. \citet{hermes13b} and \citet{gianninas14} discuss the purity of the instability strip for ELM WDs, and demonstrate that it may not be pure due to the complicated evolutionary tracks for ELM WDs that go through H shell flashes \citep[e.g.,][]{driebe98}. The other three pulsar companions (PSR J1012+5307, J1909$-$3744, and J1911$-$5958A) are close to the empirical boundaries of the instability strip, but do not vary. A new spectroscopic analysis of these pulsar companions with the same set of models as in \citet{gianninas14} will be useful to improve the empirical boundaries of the instability strip for ELM WDs, though this is outside the scope of this paper.

\subsection{Future Prospects}

There are now four ELM WD companions to milli-second pulsars that have high-speed photometry available. All four objects have temperature and surface gravity measurements that place them near the empirical boundaries of the ZZ Ceti instability strip for ELM WDs. PSR J1012+5307 and PSR J1909$-$3744 have stringent limits (0.7 per cent and 0.1 per cent, respectively) on the lack of variability, given that all ELM WDs discovered to date pulsate with amplitudes $>0.6$ per cent.

The discovery of pulsations in the WD companion to PSR J1738+0333 provides an exciting opportunity to constrain the interior structure of this ELM WD. \citet{antoniadis12} find that the physical properties of this WD are consistent with thick H envelopes. They also find that the cooling age of this WD is difficult to constrain and it is likely in the range $0.5-5$ Gyr. The main problem is that the cooling age depends on both the thickness of the surface H layer and the metallicity of the progenitor system.

Extensive follow-up observations of PSR J1738+0333's companion will be useful to determine all the periodicities in this star. Given that a precise radius measurement is also available from radio timing measurements, this pulsating ELM WD provides the perfect test-bed for a detailed pulsation study. This will likely reveal the detailed internal structure of an ELM WD pulsator, and constrain its cooling age \citep{vangrootel13}. Furthermore, these constraints can be used to obtain independent mass measurements for both the ELM WD and the pulsar.

\section*{Acknowledgements}

We gratefully acknowledge the support of the NSF and NASA under grants AST-1312678 and NNX14AF65G, respectively. J.J.H. acknowledges funding from the European Research Council under the European Union's Seventh Framework Programme (FP/2007-2013) / ERC Grant Agreement no 320964 (WDTracer).

\end{document}